\documentclass[
	a4paper, 
	10pt, 
	unnumberedsections, 
	twoside, 
]{LTJournalArticle}

\runninghead{} 

\footertext{\textit{RISC-V Summit Europe, Bologna, 8-12th June 2026}} 

\title{Accelerating Sparse Linear Solvers in OpenFOAM using RISC-V Vector Extensions} 

\author{%
	Gabriele Ceccolini\textsuperscript{1}\thanks{Corresponding author: \href{mailto:gabriele.ceccolini@studio.unibo.it}{\tt gabriele.ceccolini@studio.unibo.it}}, 
    Federico Ficarelli\textsuperscript{2},
    Filippo Barbari\textsuperscript{2},
    Simone Bnà\textsuperscript{2}
    and Andrea Bartolini\textsuperscript{1}
}

\date{\footnotesize\textsuperscript{\textbf{1}}University of Bologna, Italy \\ \textsuperscript{\textbf{2}}CINECA - HPC Department, Italy\\ }

\renewcommand{\maketitlehookd}{%
	\begin{abstract}
		\noindent 
    Computational Fluid Dynamics (CFD) relies heavily on the efficiency of linear solvers based on sparse linear algebra kernels.
    Widely used frameworks like OpenFOAM exploit parallelism primarily at the domain decomposition level via MPI.
    Support for vector/SIMD architectures is limited to compiler auto-vectorization.
    Furthermore, support for such architectures is limited by OpenFOAM's internal matrix data format, which is intrinsically ill-suited for the contiguous memory accesses required for efficient execution on vector processors.
    In this work, we focused on two very different RISC-V architectures: the prototype long-vector EPAC accelerator and the commercial short-vector CPU Sophon SG2044.
    On these platforms, we optimized the Sparse Matrix-Vector multiplication (SpMV) using RISC-V vector intrinsics and integrated it into a custom smoother, performing a runtime conversion of internal data into a vector-friendly format.
    Experimental results on the EPAC test chip show a $6\times$ speedup for the smoother; benchmarks on Monte Cimone (MCv2) cluster with the Sophon SG2044 processor achieve a $1.5\times$ smoother speedup, proving that legacy CFD codes can be effectively accelerated on both research and commercial emerging hardware.
	\end{abstract}
}

\begin{document}

\maketitle 

\section{Introduction}
OpenFOAM~\cite{foam} is a leading CFD framework that traditionally relies on MPI for process-level parallelism, neglecting the potential of modern vector architectures. This limitation derives from its native LDU-COO (Lower-Diagonal-Upper) matrix format, designed for scalar execution, lacking the contiguous memory access patterns required by vector processing units. While GPU acceleration has been tackled~\cite{SPUMA}, exploiting CPUs SIMD Instruction Set Architectures (ISA) in OpenFOAM remains an open challenge.
To partially address this, we introduce a runtime conversion of the data matrix into a vector-friendly layout, accelerating sparse linear algebra through RISC-V vector intrinsics. We specifically target the Geometric-Algebraic Multi-Grid (GAMG), the native multigrid OpenFOAM solver adopted for the solution of stiff elliptic problems. 
We evaluate this approach oon two RISC-V vector implementations based on \textit{radically} different design principles (a long-vector EPAC architecture with $VLEN=16384$ bits, and a short-vector SG2044 architecture with $VLEN=128$ bits) to assess how different vector length paradigms perform with highly memory-bound workloads.

The main contributions of this work are: (i) the manual vectorization through intrinsics and scalability analysis of the SpMV kernel; (ii) the implementation of a custom vectorized smoother as an external plugin for OpenFOAM's GAMG solver; and (iii) the end-to-end performance evaluation of the accelerated smoother on a real CFD multi-step simulation.
%
\vspace{-10pt}
\section{Implementation}
We targeted the Richardson smoother within GAMG, as it is inherently vector-friendly and accounts for $\sim$50\% of the entire simulation time. This smoother comprises two dense vector kernels and one SpMV.
All three of these kernels were manually vectorized using intrinsics, specifically targeting RVV 0.7 EPI for the EPAC and RVV 1.0 for the SG2044.
Major attention was dedicated to the SpMV kernel, as it dominates the computation time in the scalar version of the smoother. In particular, two of the most popular vector-friendly storage formats were explored: ELL and SELL-C-$ \sigma $~\cite{sell_c_sigma_paper}.
After evaluating the SpMV speedup of these formats against the scalar baseline, 
we integrated the vectorized kernels into a custom smoother, dynamically linking 
it to OpenFOAM at runtime. This plugin-based approach provides the significant 
advantage of allowing vector acceleration without modifying the core codebase.
%
\vspace{-10pt}
\section{Experimental results}
We evaluated the SpMV speedup on symmetric matrices taken from an OpenFOAM simulation of a laminar flow around a cylinder (128k to 4096k cells), comparing the vectorized SELL-C-$\sigma$ and ELL formats against the symmetric scalar LDU-COO (Coordinate List) baseline.
The results show an average speedup of around 7$\times$ for SpMV in the EPAC architecture using the SELL-C-$\sigma$ format, whereas the speedup is only 2$\times$ on the SG2044.
The SELL-C-$\sigma$ format consistently exhibited better memory access behavior than its ELL counterpart, as reflected by the percentage of VPU stalls on EPAC and the percentage of cache misses on SG2044.
As already observed for dense BLAS kernels~\cite{monte_cimone}, register grouping (LMUL=4) was effective in increasing the performance and speedup of SpMV on the SG2044, a feature that is precluded in EPAC by a design choice.

\begin{figure}[h] 
    \vspace{-0.4cm} 
    \centering
    \includegraphics[width=\columnwidth]{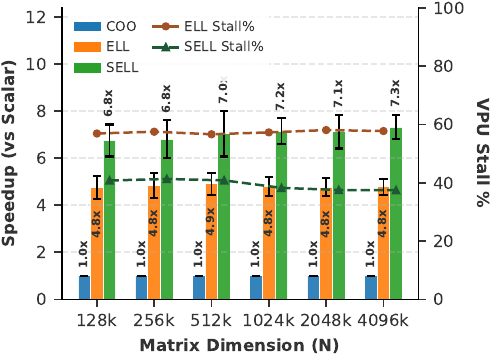}
    
    \caption{EPAC SpMV speedup and VPU stalls analysis.}
    \vspace{-0.4cm} 
    \label{fig:epac_spmv_stalls}
\end{figure}

\begin{figure}[h] 
    \vspace{-0.4cm} 

    \centering
    \includegraphics[width=\columnwidth]{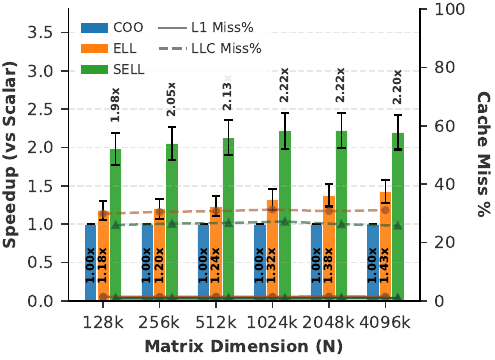}
    
    \caption{SG2044 SpMV speedup and cache miss analysis.}
    \vspace{-0.4cm} 
    \label{fig:sg2044_spmv_miss}
\end{figure}

When evaluating the end-to-end execution time of the smoother on EPAC, we observe a total speedup of $6\times$.
Part of this advantage derives from the dense, reciprocal, and Richardson vector update kernels, which were also vectorized using intrinsics.
On the SG2044, however, the overall speedup is limited to $1.5\times$; furthermore, no benefits were observed from vectorizing memory-bound dense kernels on short-vector architectures.

\begin{figure}[h] 

    \centering
    \includegraphics[width=\columnwidth]{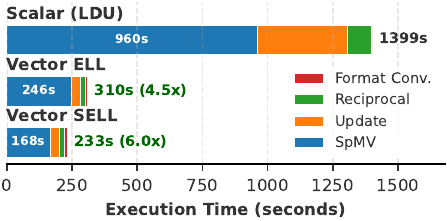}
    
    \caption{EPAC smoother profiling on 128k problem.}

    \label{fig:epac_smoother}
\end{figure}

\begin{figure}[h] 

    \centering
    \includegraphics[width=\columnwidth]{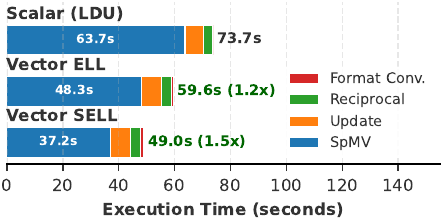}
    
    \caption{SG2044 smoother profiling on 128k problem.}
    \vspace{-0.4cm} 
    
    \label{fig:sg2044_smoother}
\end{figure}
In general, as investigated in~\cite{short_reason_for_long_vector}, long-vector architectures proved to be highly effective for memory-bound kernels, as they are able to tolerate memory latency and maximize memory bandwidth utilization.
Crucially, these architectural advantages translate into significant performance benefits specifically for non-dense workloads.
Ultimately, this work demonstrates that legacy CFD frameworks like OpenFOAM can be seamlessly accelerated via non-invasive plugins and runtime format conversions, unlocking the potential of modern RISC-V vector architectures without altering the core codebase.

\subsubsection{Acknowledgements}
This project has received funding from the European High Performance Computing Joint Undertaking (JU) under Framework Partnership Agreement No 800928 and Specific Grant Agreement No 101036168 (EPI SGA2).
The JU receives support from the European Union’s Horizon 2020 research and innovation programme and from Croatia, France, Germany, Greece, Italy, Netherlands, Portugal, Spain, Sweden, and Switzerland.
%
\vspace{-10pt}
\printbibliography 
%
%
\end{document}